\documentclass[twocolumn,preprintnumbers,prl]{revtex4}
\usepackage{amsfonts}
\usepackage[T1]{fontenc}
\usepackage{amsmath,amsbsy,amssymb,graphicx}
\usepackage{times}
\usepackage{amsmath}
\usepackage{amssymb}
\usepackage{graphicx}
\usepackage[usenames]{color}

\def\Section#1{\bigskip\noindent\textbf{\large#1}\par\noindent}
\let\section=\Section
\let\mathbf=\boldsymbol

\begin{document}

\title{{\Large Topological Phase Transition without Gap Closing}}
\author{Motohiko Ezawa$^1$, Yukio Tanaka$^2$ and Naoto Nagaosa$^{1,3}$}
\affiliation{{}$^1$ Department of Applied Physics, University of Tokyo, Hongo 7-3-1,
113-8656, Japan,}
\affiliation{{}$^2$ Department of Applied Physics, Nagoya University, Nagoya 464-8603,
Japan,}
\affiliation{{}$^3$Center for Emergent Matter Science, ASI, RIKEN, Wako 351-0198, Japan}

\begin{abstract}
Topological phase transition is accompanied with a change of topological
numbers. It has been believed that the gap closing and the breakdown of the
adiabaticity at the transition point is necessary in general. However, the
gap closing is not always needed to make the topological index ill-defined.
In this paper, we show that the states with different topological numbers
can be continuously connected \textit{without} gap closing in some cases,
where the symmetry of the system changes during the process. Here we propose
the generic principles how this is possible (impossible) by demonstrating
various examples such as 1D polyacetylene with the charge-density-wave
order, 2D silicene with the antiferromagnetic order, 2D silicene or quantum
well made of HgTe with superconducting proximity effects and 3D
superconductor Cu doped Bi$_{2}$Se$_{3}$. It is argued that such an unusual
phenomenon can occur when we detour around the gap closing point provided
the connection of the topological indices is lost along the detour path.
\end{abstract}

\maketitle



\address{{\normalsize Department of Applied Physics, University of Tokyo, Hongo
7-3-1, 113-8656, Japan }}

Topological insulator and superconductor are among the most fascinating
concepts in physics found in this decade\cite%
{Hasan,Qi,AndoReview,MB,Roy,tanakareview,Aliceareview}. It is characterized
by the topological indices such as the Chern number and the $\mathbb{Z}_{2}$
index. When there are two topological distinct phases, a topological phase
transition may occur between them. It has been generally accepted that the
gap must close at the topological phase transition point since the
topological index cannot change its quantized value without gap closing.
Note that the topological index is only defined in the gapped system and
remains unchanged for any adiabatic process. Alternatively we may think of
the edge or surface of the sample in a topological phase. Gapless edge or
surface modes appear because the boundary of the sample separates a
topological state and the vacuum whose topological indices are zero. The
phenomenon is known as the bulk-edge correspondence. We wonder if a
topological phase transition cannot occur without gap closing at all.

\begin{figure}[t]
\leftline{(a)} \vspace*{-5mm} \centerline{\includegraphics[width=0.25
\textwidth]{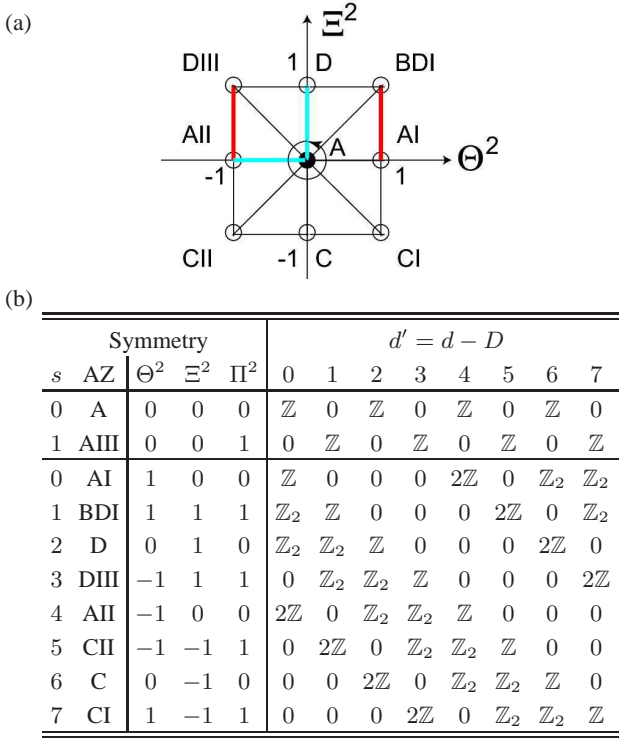}} \leftline{(b)} $
\begin{array}{cc|ccc|cccccccc}
\hline\hline
\multicolumn{5}{c@{}|}{%
\begin{array}{c}
\text{Symmetry}%
\end{array}%
} & \multicolumn{8}{c@{}}{%
\begin{array}{c}
d^{\prime }=d-D%
\end{array}%
} \\ 
s & \text{AZ} & \Theta ^{2} & \Xi ^{2} & \Pi ^{2} & 0 & 1 & 2 & 3 & 4 & 5 & 6
& 7 \\ \hline
0 & \text{A} & 0 & 0 & 0 & \mathbb{Z} & 0 & \mathbb{Z} & 0 & \mathbb{Z} & 0
& \mathbb{Z} & 0 \\ 
1 & \text{AIII} & 0 & 0 & 1 & 0 & \mathbb{Z} & 0 & \mathbb{Z} & 0 & \mathbb{Z
} & 0 & \mathbb{Z} \\ \hline
0 & \text{AI} & 1 & 0 & 0 & \mathbb{Z} & 0 & 0 & 0 & 2\mathbb{Z} & 0 & 
\mathbb{Z}_{2} & \mathbb{Z}_{2} \\ 
1 & \text{BDI} & 1 & 1 & 1 & \mathbb{Z}_{2} & \mathbb{Z} & 0 & 0 & 0 & 2
\mathbb{Z} & 0 & \mathbb{Z}_{2} \\ 
2 & \text{D} & 0 & 1 & 0 & \mathbb{Z}_{2} & \mathbb{Z}_{2} & \mathbb{Z} & 0
& 0 & 0 & 2\mathbb{Z} & 0 \\ 
3 & \text{DIII} & -1 & 1 & 1 & 0 & \mathbb{Z}_{2} & \mathbb{Z}_{2} & \mathbb{
Z} & 0 & 0 & 0 & 2\mathbb{Z} \\ 
4 & \text{AII} & -1 & 0 & 0 & 2\mathbb{Z} & 0 & \mathbb{Z}_{2} & \mathbb{Z}
_{2} & \mathbb{Z} & 0 & 0 & 0 \\ 
5 & \text{CII} & -1 & -1 & 1 & 0 & 2\mathbb{Z} & 0 & \mathbb{Z}_{2} & 
\mathbb{Z}_{2} & \mathbb{Z} & 0 & 0 \\ 
6 & \text{C} & 0 & -1 & 0 & 0 & 0 & 2\mathbb{Z} & 0 & \mathbb{Z}_{2} & 
\mathbb{Z}_{2} & \mathbb{Z} & 0 \\ 
7 & \text{CI} & 1 & -1 & 1 & 0 & 0 & 0 & 2\mathbb{Z} & 0 & \mathbb{Z}_{2} & 
\mathbb{Z}_{2} & \mathbb{Z} \\ \hline\hline
\end{array}%
$%
\caption{(a) Topological classes and possible class changes. The horizontal
axis is $\Theta ^{2}=0,\pm 1$, which represents the time-reversal symmetry,
while the vertical axis is $\Xi ^{2}=0,\pm 1$, which represents the
particle-hole symmetry. The eigenvalue $0$ means the absence of the
symmetry. We present examples of class changes by bold lines that occur
without gap closing. AIII resides with the same position as A. (b) Periodic
table for the homotopy group of each class. The rows correspond to the
different Altland Zirnbauer (AZ) symmetry classes while the columns
distinguish different dimensionalities, which depend only on $d^{\prime
}=d-D $ with $d$-dimensional $k$ space and $D$-dimensional real space
coordinates.}
\label{FigClass}
\end{figure}

The topological classes are classified\cite{Schnyder} by the eigenvalues of $
\Theta ^{2}$, $\Xi ^{2}$ and $\Pi ^{2}$, where $\Theta $, $\Xi $ and $\Pi $,
represent the time-reversal, particle-hole and chiral symmetry operators as
shown in Fig.\ref{FigClass}. There are ten classes, which are separated into
two complex and eight real representations. The topological periodic table
has been established as in Fig.\ref{FigClass}(b), which classifies all the
possible homotopy groups and topological indices depending on the symmetry
and dimensionality of the system. One important fact about this topological
periodic table is that the adiabatic connection is possible between the two
classes with the difference in dimensions by one\cite{Teo}. As for the eight
real representations, this connection is summarized by the symmetry clock
shown in Fig.\ref{FigClass}(a). Namely, considering the Hamiltonian $H(k,r)$
which depends on both the $D$-dimensional real space coordinates $r$ and $d$
-dimensional momentum space coordinates $k$, the mapping connecting the
neighboring classes in the symmetry clock is possible by adding $r$- or $k$%
-dependent Hamiltonian\cite{Teo}. This means that the essential
dimensionality is $d^{\prime }=d-D$ and the adiabatic connection exists next
to each other along the \textit{diagonal} direction in the topological
periodic table. In addition to this \textit{diagonal} shift, one can
consider the \textit{horizontal} shift, i.e., dimensionality $d^{\prime }$
by introducing the defects such as vortex ($D=1$), and point defect ($D=2)$
\cite{Reduction}. There are several works on the \textit{vertical} shift in
the periodic table\cite{QHZ,EzawaQAHE,EzawaPhoto,EzawaExM,Law}. However in
these cases, the gap closing is necessary for topological phase transitions.

In this paper, we study the adiabatic connection within the common
dimensionality $d^{\prime }$, and the possible change in the topological
indices without closing the gap. We propose two principles. Let the energy
spectrum be given by $E_{\rho }(k)$ with a topological phase transition
taking place at a critical point $\rho =\rho _{\text{cr}}$ of a certain
parameter $\rho $, where the gap closes. Let us assume that we can extend
the Hamiltonian to include a new parameter $\Delta _{s}$ so that the energy
spectrum is modified as%
\begin{equation}
E\left( k\right) =\pm \sqrt{E_{\rho }^{2}\left( k\right) +\Delta _{s}^{2}}.
\label{KeySpect}
\end{equation}%
The phase transition point is $(\rho _{\text{cr}},0)$ in the $(\rho,\Delta
_{s})$ phase diagram. We may detour the point $(\rho _{\text{cr}},0)$ in the
phase diagram, along which the gap never closes though a topological phase
transition occurs. The second principle is that the topological number
should become ill-defined by a symmetry change along the above detour. This
invalidates requirements of the gap closing when the topological number
changes.

We confirm these generic principles by demonstrating various examples. The
first example is a simple one-dimensional model of polyacetylene with
reduced symmetry at intermediate states (BDI$\rightarrow $AI$\rightarrow $
BDI) by way of the charge-density-wave (CDW). We also present a
two-dimensional example of silicene with the antiferromagnet (AF) order as
another model with symmetry reducing. We then present three models with
enhanced symmetry at intermediate state (AII$\rightarrow $DIII$\rightarrow $
AII) by way of introducing the superconducting (SC) order\cite{Schnyder}.
They are silicene and quantum well made of HgTe as two-dimensional models,
and superconductor Cu doped Bi$_{2}$Se$_{3}$ as a three-dimensional model.

\section{Polyacetylene with CDW order}

We start with presenting a well-known one-dimensional example of
polyacetylene with the CDW order from a new light. Polyacetylene belongs to
the class BDI. With including the CDW order, the class change occurs into AI
by breaking the time-reversal and particle-hole symmetries simultaneously.
We now show that there are two ways of topological phase transitions, one
(BDI$\rightarrow $BDI) with gap closing and the other (BDI$\rightarrow $AI$
\rightarrow $BDI) without gap closing.

\begin{figure}[t]
\centerline{\includegraphics[width=0.5\textwidth]{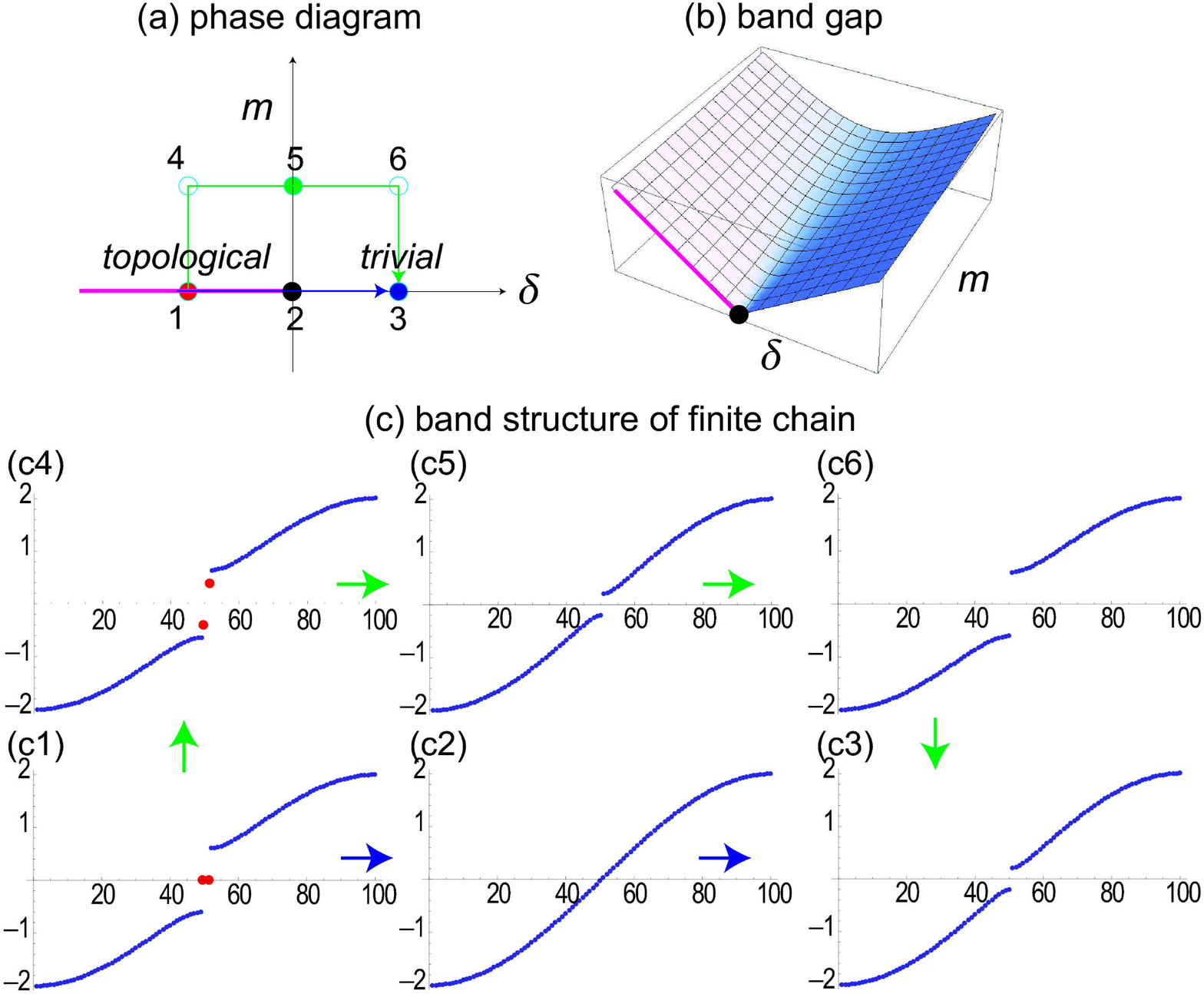}}
\caption{(a) Topological phase diagram in the $(\protect\delta ,m)$ plane.
The horizontal axis is the dimerization $\protect\delta $ and the vertical
axis is the CDW gap $m$. The band gap closes at the point denoted by a
filled circle. The system is in topological state on the red line along the $
\protect\delta $ axis. Circles show points where the energy spectrum is
calculated for finite chains in (c). (c1)$\sim $(c6) Energy spectrum of a
finite chain in each point in the phase diagram. The vertical axis is the
energy in unit of $t$ and the horizontal axis is the numbering of
eigenvalues.}
\label{FigPoly}
\end{figure}

\textbf{Effective Hamiltonian:} Polyacetylene is a bipartite system with one
unit cell made of $A$ and $B$ sites. The bipartiteness introduces a
pseudospin. We neglect the spin degree of freedom. The tight-binding model
is given by\cite{Niemi,SSH,TLM,Monceau}%
\begin{equation}
H_{\text{poly}}=d_{x}\tau _{x}+d_{y}\tau _{y}  \label{HamilPoly}
\end{equation}%
in the momentum space, where $\tau _{i}$ is the Pauli matrix acting on the
pseudospin $\Psi =\left\{ \psi _{A},\psi _{B}\right\} $, and%
\begin{equation}
d_{x}=t+\delta +\left( t-\delta \right) \cos k,\quad d_{y}=\left( t-\delta
\right) \sin k,
\end{equation}%
where $k$ is the momentum ($0\leq k<2\pi $), $t$ is the mean transfer
integral, $\delta $ is the dimerization of the transfer integral. The energy
spectrum is given by%
\begin{equation}
E_{\text{poly}}(k)=\pm 2\sqrt{t^{2}\cos ^{2}\frac{k}{2}+\delta ^{2}\sin ^{2}
\frac{k}{2}}.
\end{equation}%
The band gap locates at $k=\pi $, and is given by $2|E_{\text{poly}}\left(
\pi \right) |=2\left\vert \delta \right\vert $. We consider polyacetylene
with a finite length. We show the band structure in Figs.\ref{FigPoly}%
(c1),(c2),(c3). For $\delta <0$, the system is in the topological phase, as
is evidenced by the presence of the gapless edge states in Fig.\ref{FigPoly}
(c1). As $|\delta |$ increases, the gap decreases, and vanishes at $\delta
=0 $ as in Fig.\ref{FigPoly}(c2). For $\delta >0$, the gap opens again but
no gapless modes appear as in Fig.\ref{FigPoly}(c3): Hence, it is in the
trivial phase. Thus, the topological phase transition occurs with gap
closing at $\delta =0$.

We proceed to introduce CDW to polyacetylene. We assume it to generate the
site-energy difference $m$ between the $A$ and $B$ sites. The Hamiltonian is
modified as%
\begin{equation}
H_{\text{CDW}}=H_{\text{poly}}+m\tau _{z}.
\end{equation}%
The energy spectrum is modified to be%
\begin{equation}
E_{\text{CDW}}(k)=\pm \sqrt{\lbrack E_{\text{poly}}(k)]^{2}+m^{2}},
\label{CDWPoly}
\end{equation}%
which is of the form (\ref{KeySpect}). The gap does not close when $m\neq 0$.

\begin{figure}[t]
\centerline{\includegraphics[width=0.5\textwidth]{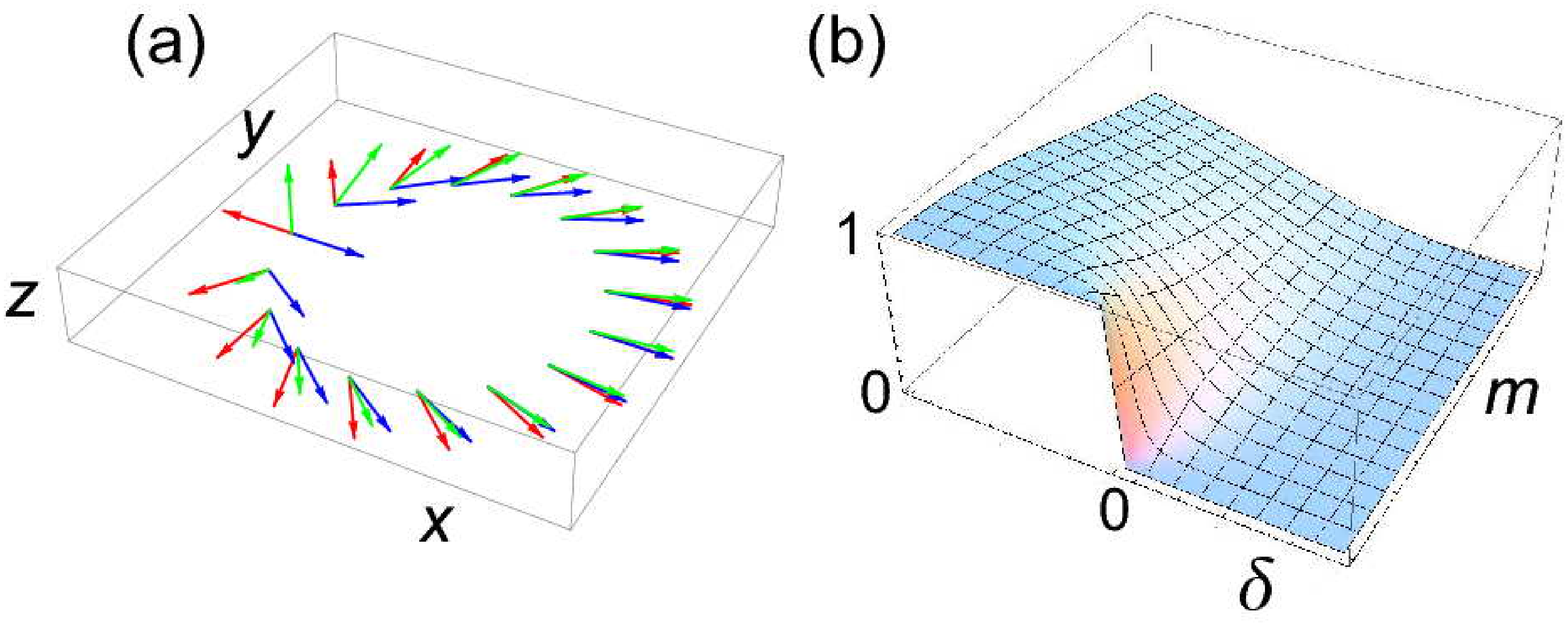}}
\caption{(a) Bipartite pseudospin in a polyacetylene chain in the momentum
space ($0\leq k\leq 2\protect\pi $). The pseudospin is confined within the $
(x,y)$ plane. The red and blue arrows correspond to the pseudospin
configurations $(n_{x},n_{y})$ for $\protect\delta <0$ and $\protect\delta 
>0 $, respectively. They belongs to different homotopy classes with the
winding number $N_{\text{wind}}=1$ and $0$. The pseudospin is allowed to
have the $z$ component in the presence of the CDW order, as indicated by the
green arrows. It connects two different classes continuously. (b) The
winding number $N_{\text{wind}}$ in the $(\protect\delta ,m)$ space. It
changes suddenly at the phase transition point $\protect\delta =0$ along the 
$\protect\delta $ axis, but smoothly when the point is detoured.}
\label{FigPolyTopo}
\end{figure}

\textbf{Phase diagram:} We explore the topological phase diagram in the $
(\delta ,m)$ plane in Fig.\ref{FigPoly}(a). We show the band gap as a
function of the electric field $\delta $ and the CDW gap $m$ in Fig.\ref
{FigPoly}(b). It is intriguing that the gapless point exists only at one
isolated point $(0,0)$ in the phase diagram. We consider two paths
connecting the topological state at $(\delta ,0)$ with $\delta <0$ and a
trivial state at $(\delta ,0)$ with $\delta >0$ shown in the phase diagram.
In Fig.\ref{FigPoly}(c) we show the energy spectrum of finite chain at
typical points.

We have already studied the first path along the $\delta $ axis. As the
second path, we first move along the $m$ axis. As changing $\delta $, there
is no gap closing even at $\delta =0$ due to the CDW gap: See Fig.\ref
{FigPoly}(c5). When $\delta $ exceeds $0$ as in Fig.\ref{FigPoly}(c6) we
remove the CDW order. The resultant phase is a trivial insulator, as is
given by Fig.\ref{FigPoly}(c3) on the $\delta $ axis. Along this process the
gap never closes. This is an explicit example of a topological phase
transition without gap closing.

\textbf{Topological analysis:} We analyze the topological number. For this
purpose we define%
\begin{eqnarray}
N_{\text{wind}}(m) &=&\frac{1}{2\pi }\int_{0}^{2\pi }dk\,[n_{x}\partial
_{k}n_{y}-n_{y}\partial _{k}n_{x}]  \notag \\
&=&\frac{1}{2}-\frac{t\delta +m^{2}}{2\sqrt{t^{2}+m^{2}}\sqrt{\delta
^{2}+m^{2}}},  \label{WindNum}
\end{eqnarray}
where $(n_{x},n_{y},n_{z})=(d_{x},d_{y},m)/\sqrt{d_{x}^{2}+d_{y}^{2}+m^{2}}$
is the normalized vector, which we illustrate in Fig.\ref{FigPolyTopo}(a).

When $m=0$, the pseudospin is confined in the $xy$ plane, and the homotopy
class is $\pi _{1}(S^{1})=\mathbb{Z}$. Correspondingly, the quantity (\ref
{WindNum}) takes only two values; $N_{\text{wind}}=1$ for $\delta <0$, and $
N_{\text{wind}}=0$ for $\delta >0$. It is the winding number, as explained
in Fig.\ref{FigPolyTopo}(a). Indeed, the system is topological for $\delta
<0 $, and trivial for $\delta >0$.

On the other hand, when the CDW is present ($m\neq 0$), the pseudospin
acquires the $z$ component, and the homotopy class changes to the trivial
class $\pi _{1}(S^{2})=0$. The quantity (\ref{WindNum}) is no longer the
quantized to be an integer. Indeed, we can continuously change $N_{\text{wind
}}(m)$ from $N_{\text{wind}}(m)=1$ to $N_{\text{wind}}(m)=0$ as we move in
the $(\delta ,m)$ plane: See Fig.\ref{FigPolyTopo}(b).

\section{\textbf{Silicene with AF order}}

We next present a two-dimensional example of silicene with the AF order as
another model with symmetry reducing. Silicene is a honeycomb structure made
of silicone atoms. It is a quantum spin-Hall (QSH) insulator\cite{LiuPRL},
and belongs to the class AII. When we introduce the AF order in the $z$
axis, the class changes into A, but it is still a topological insulator. We
call it spin-Chern insulator because the time-reversal symmetry is broken.
We show that there are two ways of topological phase transitions between the
QSH insulator and the trivial insulator with and without gap closing.

\textbf{Low-energy Dirac theory: }We analyze the physics of electrons near
the Fermi energy, which is described by Dirac electrons near the $K$ and $
K^{\prime }$ points. We also call them the $K_{\eta }$ points with the
valley index $\eta =\pm $. We introduce the AF order $m_z$ along the $z$
direction. The low-energy Dirac theory reads\cite{KaneMele,LiuPRB,EzawaExM}
\begin{equation}
H_{\eta }^{\text{AFz}}=\hbar v_{\text{F}}\left( \eta k_{x}\tau
_{x}+k_{y}\tau _{y}\right) +\eta \lambda _{\text{SO}}\sigma _{z}\tau
_{z}-m_{z}\sigma _{z}\tau _{z},  \label{DiracAF}
\end{equation}%
where $\sigma _{a}$ and $\tau _{a}$ with $a=x,y,z$ are the Pauli matrices of
the spin and the sublattice pseudospin, respectively. The first term arises
from the nearest-neighbor hopping, where $v_{\text{F}}=\frac{\sqrt{3}}{2\hbar
}at=5.5\times 10^{5}$m/s is the Fermi velocity with the lattice constant $
a=3.86$\AA . The second term is the intrinsic spin-orbit interaction with $
\lambda _{\text{SO}}=3.9$meV. The third term represents the AF order. We
have neglected the Rashba interaction since its existence does not modify
the essential part of the physics. We present the Hamiltonian containing it
in Supplementary Information.

\begin{figure}[t]
\centerline{\includegraphics[width=0.5\textwidth]{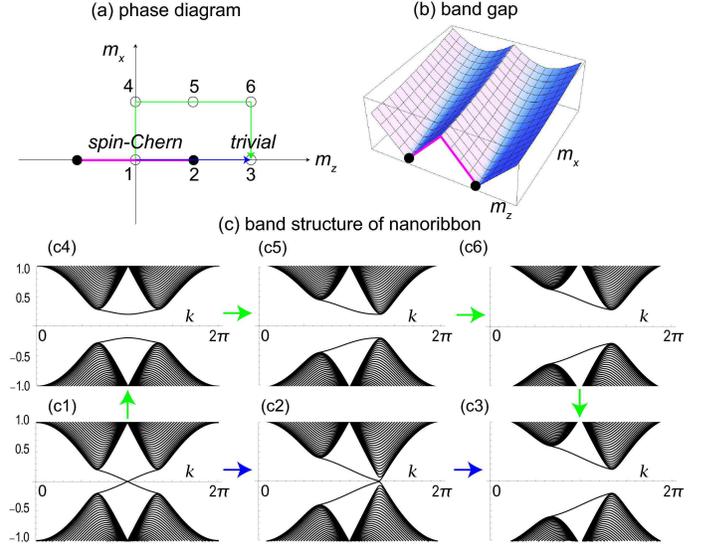}}
\caption{(a) Topological phase diagram in the $(m_{z},m_{x})$ plane. The
horizontal and vertical axes are the AF orders $m_{z}$ and $m_{x}$,
respectively. (c) We have set $\protect\lambda _{\text{SO}}=0.2t$ for
illustration. See also the caption of Fig.\protect\ref{FigPoly}.}
\label{FigInAF}
\end{figure}

The system undergoes a topological phase transition as the AF order $m_{z}$
changes\cite{EzawaExM}. The energy spectrum is given by%
\begin{equation}
E^{\text{AFz}}\left( k\right) =\pm \sqrt{\left( \hbar v_{\text{F}}\right)
^{2}k^{2}+\left( \eta \lambda _{\text{SO}}-m_{z}\right) ^{2}}.
\end{equation}%
The system is a topological insulator for $|m_{z}|<2\lambda _{SO}$ with the
gap $2|\lambda _{SO}-m_{z}|$, as is evidenced by the emergence of gapless
edge modes based on the bulk-edge correspondence: See Fig.\ref{FigInAF}(c1).
As $\eta m_{z}$ changes at the $K_{\eta }$ point, the gap decreases and
closes at the critical point $m_{z}=\eta \lambda _{\text{SO}}$, as in Fig.\ref{FigInAF}(c2). Then, the gap opens again, but there appear no longer
gapless edge modes as in Fig.\ref{FigInAF}(c3), indicating that the system
is in the trivial phase. The Chern and spin-Chern numbers are calculated to
be%
\begin{equation}
(\mathcal{C},\mathcal{C}_{\text{spin}})=\left\{ 
\begin{array}{c}
(0,1)\text{ for }|m_{z}|<\lambda _{\text{SO}} \\ 
(0,0)\text{ for }|m_{z}|>\lambda _{\text{SO}}%
\end{array}%
\right. .
\end{equation}%
This is a typical example of a topological phase transition with gap closing.

Let us now introduce the AF order additionally in the $x$ and $y$
directions. The low-energy Dirac theory is given by

\begin{equation}
H_{\eta }^{\text{AF}}=H_{\eta }^{\text{AFz}}-\left( m_{x}\sigma
_{x}+m_{y}\sigma _{y}\right) \tau _{z}.
\end{equation}%
The energy spectrum is given by%
\begin{equation}
E^{\text{AF}}\left( k\right) =\pm \sqrt{\lbrack E^{\text{AFz}}\left(
k\right) ]^{2}+m_{x}^{2}+m_{y}^{2}},
\end{equation}%
which is of the form (\ref{KeySpect}). The band gap locates at $k=0$, and is
given by $2|E_{\text{si}}\left( 0\right) |=2\sqrt{\left( \eta \lambda _{\text{SO}}-m_{z}\right) ^{2}+m_{x}^{2}+m_{y}^{2}}$. It closes only at $(m_{x},m_{y},m_{z})=(0,0,\eta \lambda _{\text{SO}})$.

\textbf{Phase diagram:} We explore the phase diagram in the $(m_{z},m_{x})$
plane with $m_{y}=0$ in Fig.\ref{FigInAF}(a). We show the band gap as a
function of the AF orders $m_{z}$ and $m_{x}$ in Fig.\ref{FigInAF}(b). The
gapless points exist only at two isolated points $(\eta \lambda _{\text{SO}
},0)$ in the phase diagram. We consider two paths connecting the QSH state
at the origin and a trivial state at $(m_{z},0)$ with $m_{z}>\lambda _{\text{
SO}}$ shown in the phase diagram. In Fig.\ref{FigInAF}(c) we show the band
structure of silicene with edges at typical points. Along the detour the gap
never closes. Hence we have shown that there are two ways of topological
phase transitions with and without gap closing.

\textbf{Topological analysis:} Silicene is a topological insulator
characterized by the Chern number and the $\mathbb{Z}_{2}$ index. It is to
be noted that the $z$ axis has been chosen by the intrinsic SO interaction
in the Hamiltonian (\ref{DiracAF}). As far as the AF order is along the $z$
axis, there exists a rotational symmetry around the $z$ axis. The system is
the sum of two decoupled systems of $s_{z}=1/2$ and $-1/2$, and their
respective Chern numbers are the topological indices in this case.
Equivalently, one can define the sum (Chern number) and the half of the
difference (spin-Chern number) of these two Chern numbers. Note that the $
\mathbb{Z}_{2} $ index is well defined when the time-reversal symmetry is
present while the spin-Chern number is well defined when the spin $s_{z}$ is
a good quantum number. They are equal mod$\,2$ when both of them are well
defined\cite{Prodan}.

When $m_{x}=m_{y}=0$, the spin-Chern number $\mathcal{C}$ is well-defined.
The topological phase is indexed by $\mathcal{C}_{\text{spin}}=1$. However,
when we introduce $m_{x}$, the AF direction cants, and the spin-Chern number
becomes ill-defined. The system becomes a trivial insulator for $m_{x}\neq 0$
or $m_{y}\neq 0$. After we make a detour around the critical point, we
decreases $m_{x}$ until $m_{x}=0$, where the spin-Chern number becomes
well-defined again. However, the system is in the trivial phase with $%
\mathcal{C}_{\text{spin}}=0$ when $m_{z}>\lambda _{\text{SO}}$.

\textbf{Helical edge states with AF order:} It is quite interesting that the
QSH insulator becomes a trivial insulator as soon as $m_{x}$ is introduced.
(We assume $m_{z}=0$ for simplicity.) We are able to construct an effective
low energy theory to explain how such a transition occurs. The helical edges
are described by the $4\times 4$-matrix Hamiltonian given by%
\begin{equation}
H_{\text{edge}}=\left( 
\begin{array}{cc}
(\lambda _{\text{SO}}/t)\hbar v_{\text{F}}k\sigma _{z} & m_{x}\sigma _{x} \\ 
m_{x}\sigma _{x} & -(\lambda _{\text{SO}}/t)\hbar v_{\text{F}}k\sigma _{z}
\end{array}
\right) .
\end{equation}%
It is diagonalized as%
\begin{equation}
E_{\text{BdG}}\left( k\right) =\sqrt{\left( \frac{\lambda _{\text{SO}}}{t}
\hbar v_{\text{F}}k\right) ^{2}+m_{x}^{2}}.  \label{HelicGap}
\end{equation}
The Hamiltonian describes two edge modes crossing at $k=0$ for $m_{x}=0$. As
soon as $m_{x}\neq 0$, the level crossing turns into the level anticrossing,
with open gap. The gap monotonously increases as $|m_{x}|$ increases. The
system is topological for $m_{x}=0$ and is trivial for $m_{x}\neq 0$.

\section{\textbf{Silicene with SC order}}

We then present three models with symmetry increasing transitions (AII$
\rightarrow $DIII) by way of superconducting proximity effect\cite%
{Fu,FuKaneSC,Alicea,Sau,Lutchyn2010,Oreg}. The system is in the class AII
without the SC order, and the gap closing occurs at the topological phase
transition point separating the two phases with $\mathbb{Z}_{2}=1$ and $0$.
However, we are able to make a topological phase transition (AII$\rightarrow 
$DIII$\rightarrow $AII) to occur without gap closing by switching on and off
the SC order.

The first model is silicene with the SC order, which might be experimentally
achieved by silicene synthesized on the Ir substrate, which has recently
been found\cite{Ir}. Its prominent feature is that Ir is superconducting at
0.1K. This opens a natural way to fabricate superconducting proximity
effects on to silicene by cooling down the system made of silicene together
with the Ir substrate.

\begin{figure}[t]
\centerline{\includegraphics[width=0.5\textwidth]{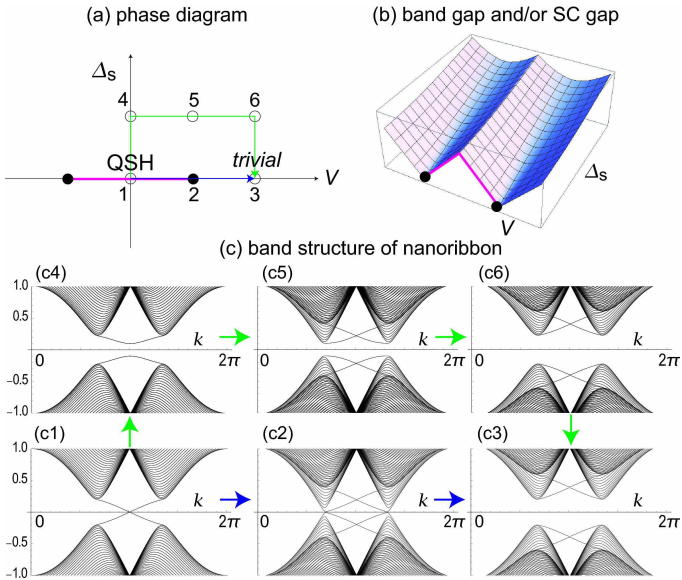}}
\caption{(a) Topological phase diagram in the $(V,\Delta _{s})$ plane. The
horizontal axis is the electric field $V$ and the vertical axis is the
superconducting gap $\Delta _{s}$. (c) We have set $\protect\lambda _{\text{
SO}}=0.2t$ for illustration. See also the caption of Fig.\protect\ref
{FigPoly}.}
\label{FigPhase}
\end{figure}

\textbf{Low-energy Dirac theory:} We analyze silicene by applying external
electric field perpendicular to the sheet. The effective Dirac Hamiltonian
in the momentum space reads\cite{KaneMele,LiuPRB,EzawaNJP}
\begin{equation}
H_{\eta }=\hbar v_{\text{F}}\left( \eta k_{x}\tau _{x}+k_{y}\tau _{y}\right)
+\lambda _{\text{SO}}\sigma _{z}\eta \tau _{z}-V \tau _{z}.
\label{HamilSilic}
\end{equation}%
The third term is the staggered potential term $V$ induced by the electric
field.

The system exhibits a topological phase transition from a QSH insulator to a
trivial insulator as $|V|$ increases\cite{EzawaNJP}. The energy spectrum is
given by%
\begin{equation}
E_{\text{si}}\left( k\right) =\pm \sqrt{\left( \hbar v_{\text{F}}\right)
^{2}k^{2}+(\eta s_{z}\lambda _{\text{SO}}-V)^{2}},
\end{equation}%
The band gap locates at $k=0$, and is given by $2|E_{\text{si}}\left(
0\right) |=2|\eta s_{z}\lambda _{\text{SO}}-V|$. It closes at $V=\pm \lambda
_{\text{SO}}$. The QSH state has gapless edge states as in Fig.\ref{FigPhase}
(c1). At $V=\lambda _{\text{SO}}$, the gap closes as in Fig.\ref{FigPhase}
(c2), and it opens for $V>\lambda _{\text{SO}}$ as in Fig.\ref{FigPhase}
(c3). The Chern and spin-Chern numbers are calculated to be%
\begin{equation}
(\mathcal{C},\mathcal{C}_{\text{spin}})=\left\{ 
\begin{array}{c}
(0,1)\text{ for }|V|<|\lambda _{\text{SO}}| \\ 
(0,0)\text{ for }|V|>|\lambda _{\text{SO}}|%
\end{array}%
\right. ,
\end{equation}%
This is a typical example of a topological phase transition with gap closing.

We assume that Cooper pairs are formed by mixing electrons at the $K$ and $
K^{\prime }$ points and condense in silicene by attaching the s-wave SC. We
present the BCS Hamiltonian and the associated BdG Hamiltonian in
Supplementary Information.

It is straightforward to diagonalize the BdG Hamiltonian,

\begin{equation}
E_{\text{BdG}}\left( k\right) =\pm \sqrt{\lbrack E_{\text{si}}\left(
k\right) ]^{2}+\Delta _{s}^{2}},  \label{BdGSilic}
\end{equation}%
which is of the form (\ref{KeySpect}). The band gap locates at $k=0$, and it
closes at $(\pm \lambda _{\text{SO}},0)$ in the $(V,\Delta _{s})$ plane.

\textbf{Phase diagram:} We explore the phase diagram in the $(V,\Delta _{s})$
plane in Fig.\ref{FigPhase}(a). We show the band gap as a function of the
electric field $V$ and the SC gap $\Delta _{s}$ in Fig.\ref{FigPhase}(b).
The gapless points exist only at two isolated points $(\pm \lambda _{\text{SO
}},0)$ in the phase diagram. We consider two paths connecting the QSH state
at the origin and a trivial state at $(V,0)$ with $V>\lambda _{\text{SO}}$
shown in the phase diagram. In Fig.\ref{FigPhase}(c) we show the band
structure of silicene with edges at typical points.

As to the second path, we first move along the $\Delta _{s}$ axis from the
origin in the phase diagram. Namely, we cool down the sample below the
critical temperature of superconductivity. As soon as the critical
temperature is passed, a topological class change occurs by adding the
particle-hole symmetry associated with the Cooper-pair condensation. The SC
gap mixes the helical edge modes, resulting in the disappearance of gapless
edge modes, and the system becomes trivial [Fig.\ref{FigPhase}(c4)]. The
mechanism how the helical edge modes disappear is precisely the same as we
demonstrated before: See Eq.(\ref{HelicGap}). After the detour we warm up
the sample. The superconductivity disappears. The resultant phase is a
trivial insulator, as is given by Fig.\ref{FigPhase}(c3) on the $V$ axis.
Along this process the gap never closes.

\begin{figure}[t]
\centerline{\includegraphics[width=0.5\textwidth]{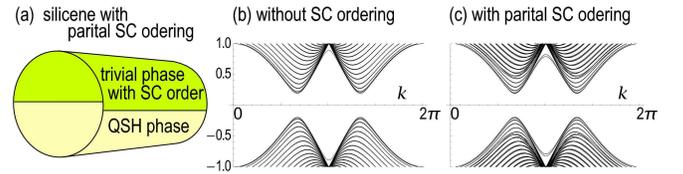}}
\caption{(a) Illustration of silicene in torus geometry with the SC order on
one-half of the system. The upper half region is in the trivial phase with
the SC order included, while the lower half is in the QSH phase. Band
structures of silicene in torus geometry (b) without the SC order and (c)
with the the SC order on one-half of the system. The vertical axis is the
energy in unit of $t$, while the horizontal axis is the momentum $k$.}
\label{FigJunction}
\end{figure}

Let us study this class change more in detail to confirm that it does not
involve the gap closing. For this purpose we analyze a system where SC is
attached on one-half of the system. It is important to demonstrate whether
there emerge edge states between the boundary of the two regions\cite
{Fu,Linder}. The region without SC is in the QSH phase, while the
superconducting region is in the trivial phase. We may alternatively
calculate the band structure of silicene in torus geometry with the SC order
introduced to one-half side of a cylinder [Fig.\ref{FigJunction}(a)]. Note
that there are no edge states in silicene in torus geometry even in the QSH
phase. The band structure is well known\cite{EzawaSiliNanotube} and given in
Fig.\ref{FigJunction}(b). Once the superconducting gap is introduced
partially as in Fig.\ref{FigJunction}(a), two boundaries appear between the
SC and normal regions. We naively expect the emergence of gapless edge modes
along each boundary due to the bulk-edge correspondence, as is the case\cite
{EzawaSiliNanotube} between the QSH phase and the trivial phase.
Nevertheless, we find no gapless edge states to appear as in Fig.\ref
{FigJunction}(c). This reflects the fact that the topological phase changes
between $\Delta _{s}=0$ and $\Delta _{s}\neq 0$ undergoes without gap
closing.

\textbf{Topological analysis:} We next search for the reason why a
topological phase transition can occur without gap closing as soon as the
superconductor gap $\Delta _{s}$ is introduced. For this purpose we
investigate the topological charges of the system. With the superconducting
proximity effects, silicene belongs to the class DIII. Thus the topological
class change occurs from AII to DIII as soon as the superconductor gap $%
\Delta _{s}$ is introduced, i.e., by adding the particle-hole and chiral
symmetries. Both classes AII and DIII are characterized by the $\mathbb{Z}%
_{2}$ indices. However they are different objects. Consequently, although we
have given the phase diagram in Fig.\ref{FigPhase}(a), each domain is
indexed by different topological indices.

Since the spin $s_{z}$ is a good quantum number along the $E$ axis in the
phase diagram, the $\mathbb{Z}_{2}$ index is essentially the spin-Chern
number for $\Delta _{s}=0$. The latter counts the numbers of up-spin
electrons and down-spin electrons separately. When the SC order is
introduced, the ground state is a condensed phase of Cooper pairs each of
which is a pair of up-spin and down spin electrons. Consequently, the spin
Chern number becomes ill-defined for $\Delta _{s}\neq 0$, and the gap
closing is not required between the QSH state at $\Delta _{s}=0$ and a
trivial state at $\Delta _{s}\neq 0$.

A comment is in order. Here we have assumed the singlet SC order. When we
assume the triplet SC order, the energy spectrum is shown to be of the form%
\begin{equation}
E_{\text{BdG}}\left( k\right) =\pm \sqrt{\left( \hbar v_{\text{F}}\right)
^{2}k^{2}+(\sqrt{\lambda _{\text{SO}}^{2}+\Delta _{t}^{2}k^{2}}-\left\vert
V\right\vert )^{2}}.
\end{equation}%
The triplet SC parameter $\Delta _{t}$ does not contribute to the gap at $%
k=0 $. The gap closing occurs although this is a topological class change
from AII to DIII together with the Bose-Einstein condensation. This is not
surprising since the spin-Chern number is well-defined for $\Delta _{t}\neq
0 $ because the members of a Cooper pair are either up-spin or down-spin
electrons.

\section{Quantum well made of HgTe and CdTe}

We present another two-dimensional example, which is the
Bernevig-Hughes-Zhang (BHZ) model of the QSH system. The BHZ model describes
the electronic structure of the subband of a quantum well made of HgTe and
CdTe\cite{BHZ}, which is experimentally verified\cite{Koenig}. A topological
phase transition can be induced by changing the thickness of the quantum
well.

\begin{figure}[t]
\centerline{\includegraphics[width=0.5\textwidth]{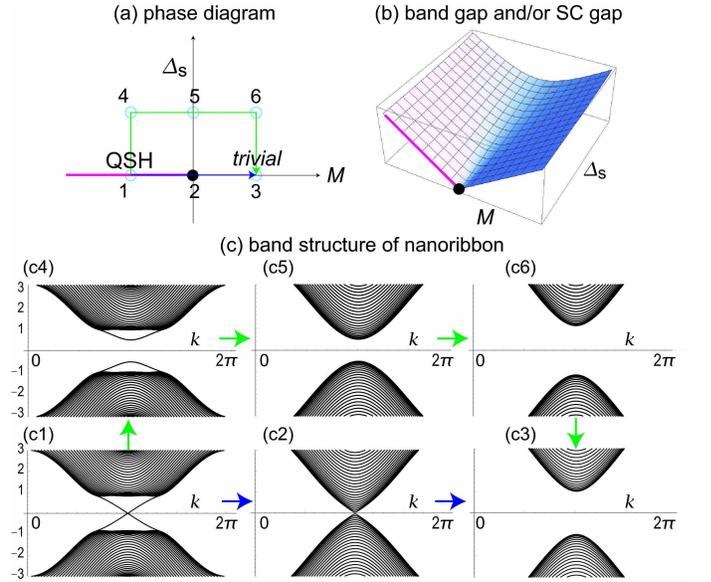}}
\caption{(a) Topological phase diagram in the $(M,\Delta _{s})$ plane. The
horizontal axis is the Dirac mass $M$ and the vertical axis is the
superconducting gap $\Delta _{s}$. The vertical axis is the energy in unit
of $B$. We have set $A=B=1,C=D=0$. See also the caption of Fig.\protect\ref%
{FigPoly}. }
\label{FigBHZ}
\end{figure}

The Hilbert space is spanned by the four states which are eigenstates of the
operator $m_{J}$, $|\pm 1/2\rangle $ and $|\pm 3/2\rangle $. We present the
BHZ Hamiltonian in Supplementary Information. The energy spectrum is
determined as%
\begin{equation}
E_{\text{BHZ}}\left( k\right) =C-Dk^{2}\pm \sqrt{Ak^{2}+(M-Bk^{2})^{2}},
\label{EneBHZ}
\end{equation}%
where $A,B,C,D$ are sample-dependent parameters, and $M$ is the Dirac mass.
The topological phase transition is known to occur at $M=0$.

To explain it let us slightly simplify the model by choosing $C=0$. Then,
the energy spectrum is simplified as%
\begin{equation}
E_{\text{BHZ}}\left( k\right) =-Dk^{2}\pm \sqrt{Ak^{2}+(M-Bk^{2})^{2}}.
\end{equation}%
The band gap locates at $k=0$, and is given by $2|E_{\text{BHZ}}\left(
0\right) |=2\left\vert M\right\vert $. We show the band structure of
silicene with edges in Figs.\ref{FigBHZ}(c1),(c2),(c3). For $M/B<0$, the
system is in the topological phase, as is evidenced by the presence of the
gapless edge modes in Fig.\ref{FigBHZ}(c1). As $|M|$ decreases, the gap
decreases, and closes at $M=0$ as in Fig.\ref{FigBHZ}(c2). For $M/B>0$, the
gap opens again but no gapless modes appear as in Fig.\ref{FigBHZ}(c3):
Hence, it is in the trivial phase. Thus, the topological phase transition
occurs with gap closing.

We proceed to assume that the singlet Cooper pairs are formed due to the SC
proximity effects with the SC gap $\Delta _{s}$. The first pairing is
between $|1/2\rangle $ and $|-1/2\rangle $. The second pairing is between $%
|3/2\rangle $ and $|-3/2\rangle $. In general these two gaps are different,
but we assume them to be equal for simplicity. The BdG Hamiltonian is
derived and given in Supplementary Information. It is diagonalized as%
\begin{equation}
E_{\text{BdG}}\left( k\right) =\pm \sqrt{\left( E_{\text{BHZ}}\left(
k\right) \right) ^{2}+\Delta _{s}^{2}},  \label{BHZSC}
\end{equation}%
where each level is two-fold degenerate. The gap does not close when $\Delta
_{s}\neq 0$.

We show the topological phase diagram in the $(M,\Delta _{s})$ plane, the
band gap and the band structure of silicene with edges with straight edge in
Fig.\ref{FigBHZ}. Employing the same discussion as in the case of silicene,
we find that it is possible to make a topological phase transition without
gap closing.

\section{Cu doped Bi$_{2}$Se$_{3}$}

Finally we present a three-dimensional example. Bi$_{2}$Se$_{3}$ is a
three-dimensional topological insulator\cite{K1}. On the other hand, Cu
doped Bi$_{2}$Se$_{3}$\cite{Hor,Wray,FuBerg} becomes superconducting at low
temperature\cite{Hor,Wray,FuBerg,AndoY}. The topological class change occurs
from AII to DIII.

The Hamiltonian is given by\cite{Hashimoto,YamakageR} 
\begin{equation}
H_{\text{BiSe}}=m(k)\tau _{x}+v_{z}k_{z}\tau _{y}+v\tau _{z}\left(
k_{x}\sigma _{y}-k_{y}\sigma _{x}\right) -\mu ,
\end{equation}%
with the chemical potential $\mu $, and%
\begin{equation}
m(k)=m_{0}+m_{1}k_{z}^{2}+m_{2}\left( k_{x}^{2}+k_{y}^{2}\right) ,
\end{equation}%
where $\sigma _{a}$ and $\tau _{a}$ with $a=x,y,z$ are the Pauli matrices of
the spin and the orbital pseudospin, respectively: $(v,v,v_{z})$ is the
Fermi velocity, and $m_{i}$ are sample parameters. The system is topological
when $m_{0}m_{1}<0$ and trivial when $m_{0}m_{1}>0$. The energy spectrum is
given by%
\begin{equation}
E_{\text{BiSe}}\left( k\right) =-\mu \pm \sqrt{%
v^{2}(k_{x}^{2}+k_{y}^{2})+v_{z}^{2}k_{z}^{2}+m(k)^{2}}.
\end{equation}%
The band gap locates at the $\Gamma $ point with $k_{x}=k_{y}=k_{z}=0$,
where $E_{\text{BiSe}}\left( 0\right) =\pm |m_{0}|$. The band gap is $%
2|m_{0}|$.

We cool the system below the SC transition point. There are four possible
superconducting pairing\cite{FuBerg}. The simplest one is the intra-orbital
spin-singlet pairing, which results in trivial superconductor. The other
three pairings lead to topological superconductor. We concentrate on the
intra-orbital spin-singlet pairing since only this pairing enables the
topological class change without gap closing. The BdG Hamiltonian is derived
and given in Supplementary Information. It is diagonalized as%
\begin{equation}
E_{\text{BdG}}\left( k\right) =\pm \sqrt{\lbrack E_{\text{BiSe}}\left(
k\right) ]^{2}+\Delta _{s}^{2}},  \label{BiSeBdGX}
\end{equation}%
where each level is two-fold degenerate. The gap does not close when $\Delta
_{s}\neq 0$. In the same way, the topological class change occurs without
gap closing in the three-dimensional space.

\section{Topological superconductor}

We have so far studied a model Hamiltonian where the initial and final
states are normal state without SC order. Here, we consider a situation
where initial and final states are SC states. We consider a model
Hamiltonian of two-dimensional spin-triplet $p_{x}$-wave superconductor with
opposite spin pairing $S_{z}=0$ or spin-singlet $d_{xy}$-wave one. Although
the initial Hamiltonian is $4\times 4$ matrix in the electron-hole and spin
spaces, it can be block diagonalized by $2\times 2$ matrix in the absence of
magnetic scattering and spin-orbit coupling. 
In order to consider the edge state for flat surface parallel to the $y$
direction, we fix momentum $k_{y}$. Then, the original two-dimensional
Hamiltonian is reduced to be one-dimensional one. The resulting Hamiltonian
can be written as 
\begin{equation}
H=\varepsilon \tau _{z}+\Delta (k_{x})\sin k_{x}\tau _{x}  \label{HamilTS}
\end{equation}%
with $\Delta (k_{x})=\Delta _{0}$ for $p_{x}$-wave pairing and $\Delta
(k_{x})=\Delta _{0}\sin k_{x}$ for $d_{xy}$-wave one, respectively.

The energy spectrum of this Hamiltonian is 
\begin{equation}
E=\pm \sqrt{\varepsilon ^{2}+\Delta ^{2}(k_{x})\sin ^{2}k_{x}}
\end{equation}%
with $\varepsilon =-t\cos k_{x}-\mu $ and $t>0$ with $\mu =\mu _{0}+t\cos
k_{y}$. 
For $|\mu |<t$, the superconducting state becomes topological with the
zero-energy surface Andreev bound state\cite{Hu,Tanaka}. On the other hand,
for $|\mu |>t$, it becomes trivial without zero energy ABS. For $|\mu |=t$,
the gap closing occurs at $k_{x}=0$ or $k_{x}=\pm \pi $. It has been
clarified that the zero energy surface Andreev bound state is protected by
the bulk topological number\cite{Sato}.

Now, let us introduce an additional term $\Delta _{1}\tau _{y}$. Then, the
time reversal symmetry is broken and the resulting Hamiltonian becomes 
\begin{equation}
H=\varepsilon \tau _{z}+\Delta (k_{x})\sin k_{x}\tau _{x}+\Delta _{1}\tau
_{y}.
\end{equation}%
The energy spectrum is obtained as 
\begin{equation}
E=\pm \sqrt{\varepsilon ^{2}+\Delta ^{2}(k_{x})\sin ^{2}k_{x}+\Delta _{1}^{2}%
}.
\end{equation}%
There is no gap closing for all $k_{x}$ and $\mu $.

We first start from the topological phase with $\Delta _{1}=0$ with $|\mu |<t
$. Next, we switch on $\Delta _{1}$. Then, the system becomes a fully gapped
one with time reversal symmetry breaking. We can change $\mu $ from $|\mu |<t
$ to $|\mu |>t$. During this change, there is no gap closing. After we
switch off $\Delta _{1}$, we reach the topological trivial phase of the
original Hamiltonian. One of the example relevant to the intermediate state
is $d_{xy}$-wave superconductor with $s$-wave pairing, where the relative
phase between these two states is $\pi /2$. The ($d_{xy}\text{+}is$)-wave
pairing was focused on as the possible surface state of $d_{xy}$-wave
pairing in the context of Cuprate\cite{Matsumoto}.

\section{Conclusions}

A topological phase transition requires in general the gap closing and the
breakdown of the adiabaticity at the transition point. This is not
necessarily the case provided a topological class or symmetry change occurs
such that the original set of topological indices become ill-defined. There
exists two possibilities of symmetry change. Both the cases with reduced
symmetry and enhanced symmetry have been considered. When the symmetry is
reduced, the target space of the Hamiltonian becomes wider, which enables us
to connect two distinct spaces adiabatically. As such an example we have
considered a one-dimensional example of polyacetylene by way of the CDW,
which induces the class change BDI$\rightarrow $AI. Here, the homotopy class
changes from $\pi _{1}(S^{1})=\mathbb{Z}$ to $\pi _{1}(S^{2})=0$, and the
winding number becomes ill-defined. Consequently, we are able to make a
topological phase transition (BDI$\rightarrow $AI$\rightarrow $BDI) to occur
without gap closing by switching on and off the CDW. We have also analyzed
silicene with the AF order as another example with symmetry reducing. It is
interesting that a topological phase transition without gap closing takes
place within the same class A. However, there is a symmetry change whether
the spin $s_{z} $ is a good quantum number and not.

When the symmetry is enhanced, the above reasoning no longer follows. As
such examples we have considered symmetry increasing transitions (AII$%
\rightarrow $DIII) by way of introducing the SC order. Although both the
classes AII and DIII are characterized by the $\mathbb{Z}_{2}$ index, their
physical meaning is different. The $\mathbb{Z}_{2}$ index is essentially the
spin-Chern number in the class AII system, which becomes ill-defined as soon
as the system moves into the SC phase (DIII) due to the SC order. We have
explicitly shown that a topological phase transition (AII$\rightarrow $DIII$%
\rightarrow $AII) may occur without gap closing by switching on and off the
SC order.

To conclude, we mention the implications of the adiabatic connection between
the different sectors of topological number. Up to now only the
single-particle Hamiltonian has been considered, but in reality the
electron-electron interaction is effective, and the gaps are often induced
by the order parameters. Therefore, the character of the quantum critical
phenomenon depends strongly on whether the continuous change of the
topological number is possible or not. Namely, when we write down the
effective action for the quantum phase transition, there are multiple order
parameters relevant to the gapless point when the continuous detour is
possible to change the topological number. When the multiple topological
phase transitions merges at one point, even more interesting quantum
critical phenomenon is expected. The global view of the phase diagram taking
into account all the possible classes will be an important direction to
study the topological quantum transition.

\section{Acknowledgements}

This work was supported in part by Grants-in-Aid (No. 22740196) for
Scientific Research from the Ministry of Education, Science, Sports and
Culture of Japan, the Funding Program for World-Leading Innovative RD on
Science and Technology (FIRST Program) and by the \textquotedblleft
Topological Quantum Phenomena\textquotedblright\ Grant-in Aid (No. 22103005)
for Scientific Research on Innovative Areas from the Ministry of Education,
Culture, Sports, Science and Technology (MEXT) of Japan.

\section{Additional information}

\textbf{Competing financial interests:} The authors declare no competing
financial interests.

\textbf{Author contributions:} ME performed the calculations. All authors
wrote the manuscript.

\end{document}